\documentclass[fdp,fleqn]{w-art}

\usepackage{times}
\usepackage{w-thm}

\usepackage[]{graphicx}
\usepackage[]{pdfsync, amsmath,amsfonts}
\begin{document}
\newcommand{\A}{\mathcal{A}}
\newcommand{\hh}{\mathcal{H}}
\newcommand{\C}{\mathbb{C}}
\newcommand{\I}{\mathbb{I}}
 \def\ds{\partial\!\!\!\slash}
 \def\M{{\mathcal M}}



\title[Higgs NCG ]{Higgs mass in Noncommutative Geometry}


\author[Devastato]{A. Devastato\inst{1,2}
  }
\address[\inst{1}]{Dipartimento di Fisica, Universit\`{a} di Napoli {\sl Federico II}}
\address[\inst{2}]{INFN, Sezione di Napoli}
\address[\inst{3}]{Departament de Estructura i Constituents de la Mat\`eria, Universitat de Barcelona}
\author[Lizzi]{F. Lizzi\inst{1,2,3}
}
\author[Martinetti]{P. Martinetti\inst{1,2}
\footnotetext{%
agostino.devastato@na.infn.it, fedele.lizzi@na.infn.it, martinetti@na.infn.it}, 
%
}

\begin{abstract}
In the noncommutative geometry approach to the standard model, an
extra scalar field $\sigma$ - initially suggested by particle physicist to
stabilize the electroweak vacuum - makes the computation of the Higgs
mass compatible with the $126$ GeV experimental value. We give a
brief account on how to generate this field from the Majorana mass of
the neutrino, following the principles
of noncommutative geometry.
\medskip

\noindent \emph{Talk given by  P.M. at Corfou Workshop on
  noncommutative field theory and gravity, september 2013}
\end{abstract}
\maketitle
\vspace{-.5truecm}
The idea that the Higgs field is somehow related to a noncommutativity
of spacetime emerged in the late 80'- early 90'\cite{Connes:1990ix,Dubois-Violette:1989fk}
and reached its full achievement with Connes and al 
\cite{Chamseddine:1996kx,Connes:1996fu} description of the Standard Model
of elementary particles [SM] in terms of \emph{noncommutative
geometry} [NCG]. The latter is a generalization of Riemannian geometry,
that allows to incorporate in a single geometrical object the
gravitational degrees of freedom (the commutative
part of the geometry) and the quantum ones (the noncommutative
part). The central object in this description is a
generalized Dirac operator $D$, whose components are the usual
Dirac matrices, the Yukawa couplings of the fermions and the mixing
parameters for quarks and neutrinos. The SM Lagrangian minimally coupled to the Einstein-Hilbert action (in
Euclidean signature) is retrieved from one single action formula. Furthermore, the mass of the Higgs boson comes as a function
of the other parameters of the theory, thus can be calculated. The
model has been through various improvements \cite{Chamseddine:2007oz},
but the prediction was always 
 around $m_H=170$ GeV, a value ruled out by Tevratron in 2008.

\begin{figure}[hbt ]
\mbox{\rotatebox{0}
{\scalebox{.6}{\includegraphics{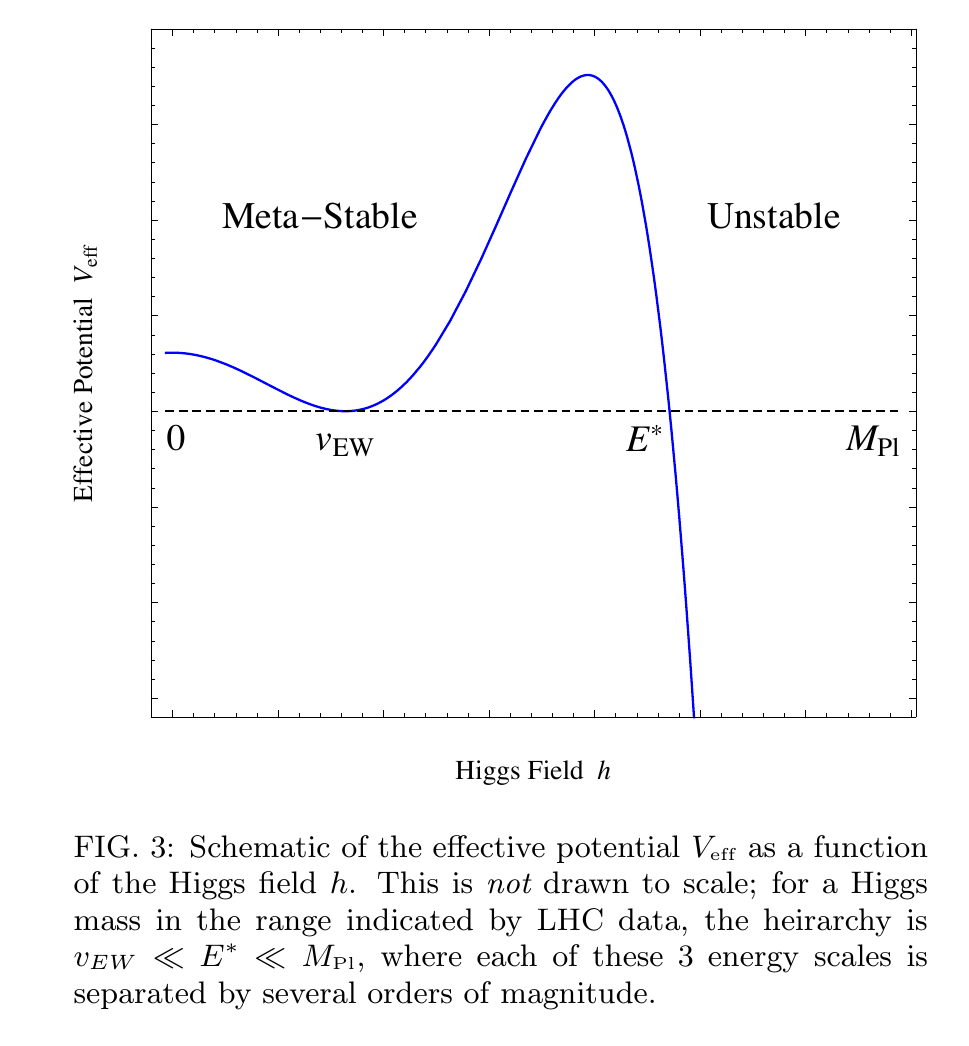}}}}
\mbox{\rotatebox{0}{\scalebox{.6}
{\includegraphics{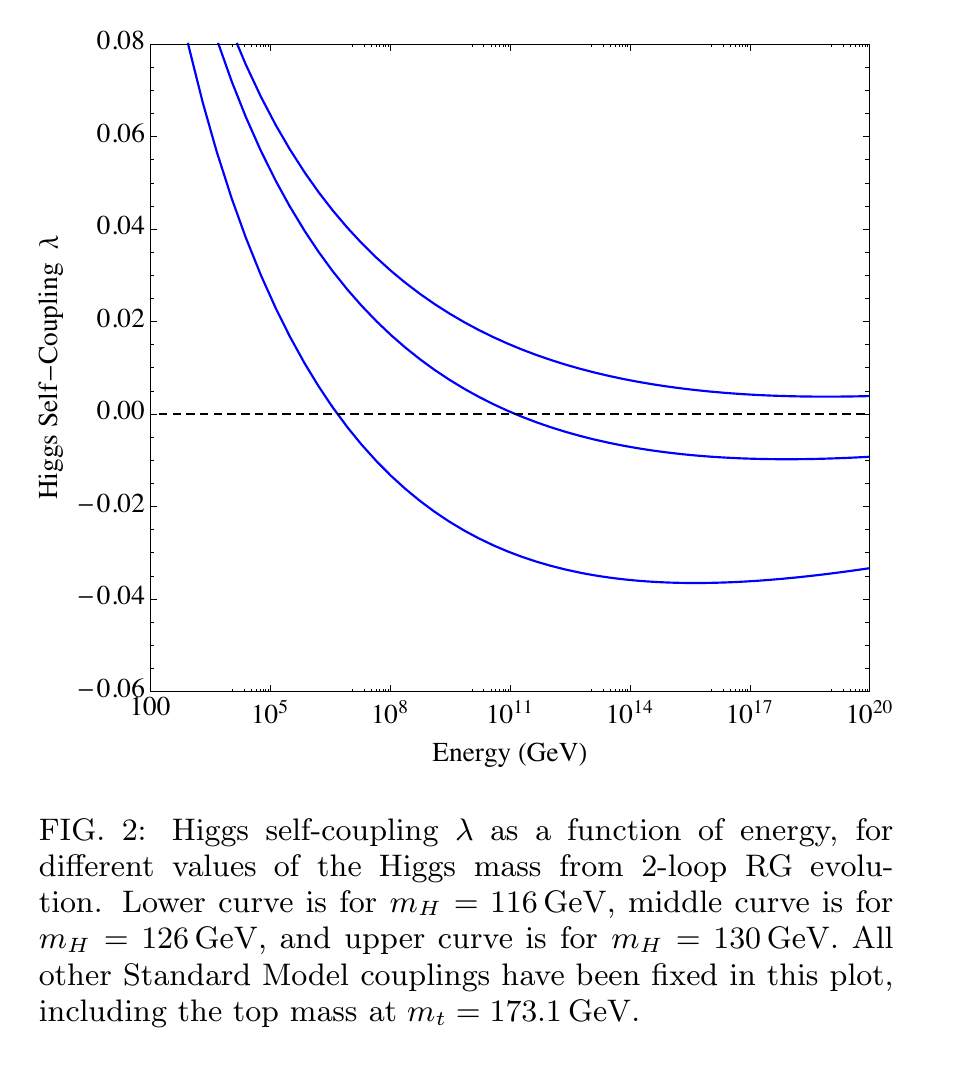}}}}
\caption{
\label{potential}
Higgs quartic selfcoupling and the effective potential (from \cite{Hertzberg:2012fk}).}
\end{figure}

\begin{figure}[h]
\vspace{-.5truecm}\mbox{\rotatebox{0}{\scalebox{.75}
{\includegraphics{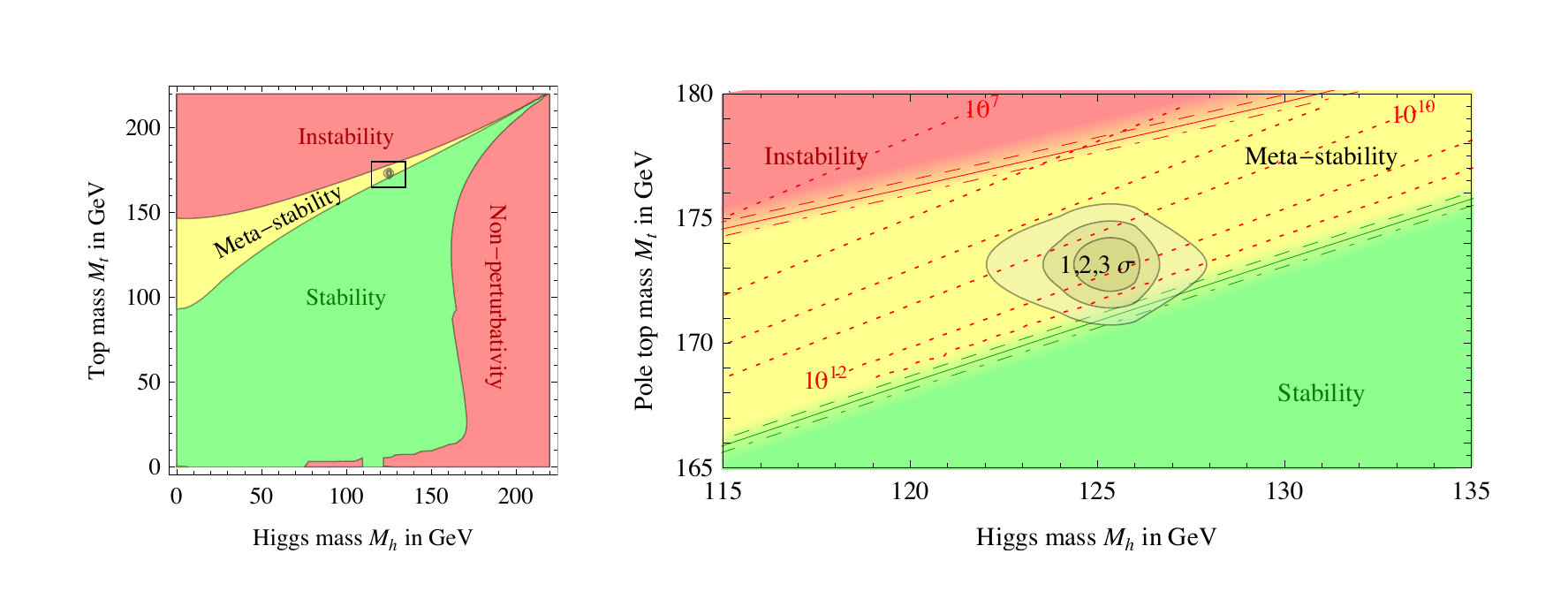}}}}
\vspace{-.6truecm}
\caption{\label{stability} Stability of the electroweak vacuum in the
  $m_{\text{top}}, m_H$ plane (from \cite{Degrassi:2012fk}).}
\end{figure}

The discovery of a $126$ GeV Higgs boson in summer 2012 at LHC
raised the question of the stability of the electroweak vacuum: the
quartic selfcoupling in the Higgs potential 
becomes negative at high energy, indicating the vacuum is
not stable but metastable (fig. \ref{potential}). There does not seem
to be a consensus whether this is a real problem or not: on the one
hand the
lifetime of this metastable state is longer than the age of the universe,  on the
other hand metastability may have some
cosmological consequence, it seems unlikely that at early age the
Higgs field has been trapped everywhere in the false vacuum, and in some
firewall scenario the metastability might even have catastrophic
consequences  (there is a vast literature on the subject, see e.g. \cite{Degrassi:2012fk} and
references within for a recent account). Furthermore, 
$126$ Gev is very close to the stability zone (fig. \ref{stability}), suggesting that
new physics may be around the corner.
This instability can be cured by a new scalar field
$\sigma$ coupled to the Higgs (e.g. \cite{Elias-Miro:2012ys}).
As a bonus, by taking into account this field  in the
description of the SM in NCG, it is possible to bring the value of $m_H$ to
$126\text{ GeV}$, without modifying the fermion
contain of the SM \cite{Chamseddine:2012fk} (using an extra
scalar field in NCG to lower $m_H$ was already in \cite{Stephan:2009fk}, but required new fermions). The point is thus to understand how to obtain
$\sigma$ intrinsically within the NCG framework. 
\smallskip

In  \cite{Chamseddine:2012fk} this is done by turning into a
field the (constant) Majorana mass term $k_R$ of the neutrino, which is one of the
component of the generalized Dirac operator $D$.
However the substitution
$k_R \rightarrow \sigma k_R$
is somehow 
ad-hoc: why should this - and only this - component of $D$ become a
field? In this proceeding we give a
non-technical summary of our recent proposal \cite{Devastato:2013fk} on how to obtain $\sigma$
within the NCG framework. We also discuss our result, in
particular in the light of another proposal made almost simultaneously
in \cite{Chamseddine:2013uq,Chamseddine:2013fk}.\newpage

A \emph{spectral triple} is a  $*$-algebra $\A$ that acts faithfully
on a Hilbert space $\hh$, together with an operator $D$ on $\hh$ such that $[D, a]$ is bounded and  $a[D - \kappa{{\mathbb{I}}}]^{-1}$ is compact
for all $a\in \A$ and $\kappa\notin \text{ Sp } D$. 
 With some extra-conditions that are the algebraic version of the
 geometrical properties of a Riemannian manifold, and that include the
definition of two more operators (a chirality $\Gamma$ and a real
structure $J$), Connes showed \cite{connesreconstruct} that to any
spectral triple $(\A, \hh, D)$ with $\A$ unital and commutative is
associated a compact Riemannian spin manifold $\M$ such that
$\A=C^\infty(\M)$. These conditions easily adapt to the
noncommutative case. The ones that are important for the present work
are the grading, the order $0$ and the order $1$ conditions:
\begin{equation}
  \label{eq:1}
  [\Gamma, a]= 0, \qquad [a, JbJ^{-1}]=0,\qquad [[D, a],  JbJ^{-1}]=0 \quad
  \forall a, b\in\A.
\end{equation}

The spectral triple of the SM is the product of the spectral triple
$(C^\infty(\M), L^2(\M,S), \ds)$ of a compact Riemannian manifold $\M$
(here $L^2(\M,S)$ is the Hilbert space of spinors and $\ds$ is the usual Dirac
operator) by a finite dimensional spectral triple 
\begin{align}
  \label{eq:2}
  \A_{sm} = \C \oplus {\mathbb{H}} \oplus M_3(\C), \quad \;
  \hh_F= \C^{N\times 32} =\mathcal H_R \oplus \mathcal
  H_L\oplus \oplus \mathcal H_R^c \oplus \mathcal H_L^C&\\[5pt]
D_F=\left(\begin{array}{cccc}
0_{8N} & M &M_R&0_{8N}\\
M^\dagger & 0_{8N} & 0_{8N} &0_{8N}\\
M_R^\dagger &0_{8N} & 0_{8N} &\bar M\\
0_{8N} &0_{8N} & M^T&0_{8N}\end{array}\right) \quad N=\sharp \text{ generations}.
\end{align}
The Hilbert space $\hh_R= \C^{N\times 8}$ is the space of right fermions ($6$ colored
quarks + $1$ lepton + $1$ neutrino), $\hh_L$ is the space of left fermions and $\hh_R^C,
\hh_L^c$ are the ones of antifermions. The matrix $M$ contains the quarks, leptons, neutrinos (Dirac)
mass with CKM mixing; $M_R$ contains the Majorana neutrinos mass $k_R$. 
The chirality  is $\gamma_F =\text{diag}(\mathbb I_{8N}, -\mathbb I_{8N}, -\mathbb I_{8N} , \mathbb I_{8N})$ 
and the real structure is an antidiagonal matrix with entries
$\mathbb{I}_{16N}, \mathbb I_{16N}$.
  The total spectral triple is 
\begin{equation}
\label{eq:2bis}
  {\A} = C^\infty(\M) \otimes \A_{sm},\quad \hh = L^2(\M, S)\otimes \hh_F, \quad  { D}= \ds\otimes
  \I_{N\times 32} + \gamma^5\otimes D_F
\end{equation}
with ${\Gamma} =\gamma^5\otimes \gamma_F$ and  $J =
\mathcal{J}\otimes J_F$, where $\mathcal{J}$ is the charge conjugation
operator.
\smallskip

The gauge fields of the SM (including the Higgs) are obtained by
fluctuation of the Dirac operator substituting $D$ with
\begin{equation}
D_A = D + A +
JAJ^{-1}
\label{eq:3}
\end{equation}
where $A$ is a generalized gauge potential, i.e.\ a
selfadjoint element of the form  $\sum_i a_i [D, b_i]$ with $a_i,
b_i\in \A$. 
This amounts to turning the constant components of $M$ in $D_F$ into fields on
the manifold~$\M$.  The spectral action $\text{Tr} f(\frac{D_A}{\Lambda})$
with $f$ a smooth approximation of the characteristic function of the
interval $[0,1]$ and $\Lambda$ an energy scale, then yields the bosonic part of the Lagrangian of the SM minimally coupled with
  Einstein-Hilbert action.
It requires a unique unification scale $\Lambda$. With $\Lambda =
10^{17} \text{GeV}$, the running of the Higgs quartic selfcoupling
$\lambda$ under the
big desert hypothesis yields 
$m_H =\sqrt{2\lambda}\,v_{EW}\simeq 170 \text{ GeV}.$
After having turned into a field $k_R\sigma$ the neutrino Majorana mass $k_R$,  the same procedure allows to pull back $m_H$ to 
$126\text{ GeV}.$ 

One may think to obtain $\sigma$ as the
other bosonic fields, i.e. thanks to a fluctuation of the
metric. Unfortunately the order 1 condition prevents this. Denoting $D_R$ the part of $D_F$ that contains
only the neutrino mass, one has
\begin{equation}
  \label{eq:5}
   {[[D_R,a], JbJ^{-1}] = 0 \quad\forall a, b\in \A_{sm}\Longrightarrow [D_R, a]= 0.}
\end{equation}
In other terms, because of the first order condition there is no way to obtain $\sigma$ by a fluctuation of
the Majorana part of the finite dimensional Dirac operator $D_F$.
\bigskip

In \cite{Devastato:2013fk} we proposed to obtain $\sigma$
starting from a bigger algebra than the one of the SM. 
Under natural assumptions (irreducibility of the representation,
existence of a cyclic vector), technical requirements of the NCG model
(there is a representation of the opposite algebra that commutes with
the action of the algebra and is implemented by an operator that
commutes with the chirality) and a hypothesis on the role of
quartenion, one has that the most general finite dimensional
algebra satisfying the axiom of noncommutative manifolds is of the form \cite{Chamseddine:2008uq}: 
${M_a(\mathbb H) \oplus M_{2a}(\C)},\; a\in\mathbb N$,
and acts on an Hilbert space of dimension  {$d= 2\times (2\times a)^2$}.
  The case $a=1$ is too small to get the gauge group of the SM as the
  group of unitaries of $M(\mathbb H)\oplus M_2(\C)$. 
The next choice $a=2$ yields  $d= 32$, that is the number of fermions per
    generation. As explained in \cite{Chamseddine:2008uq}, 
the grading condition imposes the reduction
\begin{equation}
  \label{eq:7}
M_2(\mathbb H) \oplus M_{4}(\C) \longrightarrow \A_{LR} = \mathbb H_L \oplus
\mathbb H_R \oplus M_4(\C). 
\end{equation}
The order 1 condition and  neutrino mass further imposes
\begin{equation}
  \label{eq:7}
\A_{LR} \longrightarrow \mathbb H_L \oplus \mathbb H_R
\oplus M_3(\C) \oplus \C  \longrightarrow
\mathbb H_L \oplus \C'
\oplus M_3(\C) \oplus \C\end{equation}
 with $\C = \C'$. Hence the
the reduction of $\A_F$ to the algebra $\A_{sm}$ of the standard model.

The case $a=3$ yields $d= 72$. There is no obvious relation with
 the $32$ particles/generation of the SM.
 Interestingly, $a=4$ yields $d= 128$, which is $4$ times 
  the number of particles/generation. Viewing $4$ as the
  number of components of a Dirac spinor on a $4$-dimensional
  manifold, one can thus decompose the total Hilbert space (for
  $1$~generation), using the fermion doubling of the model~\cite{fermiondoubling}, as
\begin{equation}
  \label{eq:9}
   {\hh = L^2(\M, S) \otimes \hh_F = L^2(\M)\otimes \mathsf H_F}
\quad\text{ where }\quad \mathsf H_F = \C^4  \otimes \hh_F = \C^4 \otimes \C^{32}=
\C^{128}.
\end{equation}
In writing \eqref{eq:9}  we ignore global obstruction and assume that
the r.h.s. equality of the first equation above holds on a local trivialization of the
spin bundle. The idea we want to promote is that by mixing the spin
degrees of freedom ($s=l,r$ for the left, right components of a Dirac
spinor, $\dot s= \dot 0, \dot 1$ for the (anti)-particles ones) with
the internal degrees of freedom
\begin{align}
C= p,a \quad (\text{particle, antiparticles}),\qquad
I=0,1,2,3\quad \text{(lepto-color)},\\
\alpha = u_R, d_R, u_L, d_L \;(I=1,2,3),   \; e_R, \nu_R, e_L,\nu_L\;
(I= 0)\quad \text{ (flavor)},
\label{eq:4}
\end{align}
then the Hilbert space $\hh$ of
the standard model is big enough to represent the  {\it grand algebra} ($a=4$)
\begin{equation}
\A_G = M_4(\mathbb H)
  \oplus M_8(\C)
\label{eq:6}
\end{equation}
(tensorized by $C^\infty(\M)$) without touching the SM particle
contents, and satisfying the order~$0$~condition.

Explicitly the representation is as follows. We denote a  spinor in
$\hh$ as $\Psi_{s\dot s\alpha}^{CI}$ and view both $ {Q\in
 M_4(\mathbb H)}$ and $ {M\in M_8(\mathbb C)}$ as $2\times 2$
 block matrices, with block $4\times 4$ complex matrices:
 \begin{equation}
 Q= \left(\begin{array}{cc}
 Q_{\dot 0\alpha}^{\dot 0\beta}& Q_{\dot 0\alpha}^{\dot 1\beta}\\
 Q_{\dot 1\alpha}^{\dot 0\beta}& Q_{\dot 1\alpha}^{\dot 1\beta}\end{array}\right),\quad 
 M = \left(\begin{array}{cc}
 M_{r J}^{r I}& M_{r  J}^{l I}\\
 M_{l J}^{r I}& M_{l  J}^{l I}\end{array}\right).
 \label{eq:17}
 \end{equation}
 Viewing the components of the matrices in \eqref{eq:17} as functions
 on $\M$, an element $A=(Q,M)$ in $C^\infty(\M)\otimes \A_G$ acts on
 $\hh$  as
  \begin{equation}
   {A_{s\dot s DJ\alpha}^{t\dot t CI\beta} =
  \left(
 \delta_0^C \delta^t_s \delta_J^I Q_{\dot s \alpha}^{\dot t \beta} +
 \delta_1^C M_{sJ}^{tI}\delta_{\dot s}^{\dot t}\delta_\alpha^\beta\right)}.
  \label{eq:18}
  \end{equation}
The indices $\beta, J$ run on the same set as $\alpha, I$.
The Dirac operator, chirality and real structure are unchanged.
The  grading condition imposes the reduction 
 \begin{equation*}
   \label{eq:13}
   \A_G = M_4(\mathbb H) \oplus M_8(\C) \longrightarrow \A'_G = (M_2(\mathbb
   H)_L \oplus M_2(\mathbb H)_R) \oplus  (M_4(\C)_l \oplus M_4(\C)_r).
\end{equation*}
  
A solution of the  {first-order condition} of the
   {Majorana Dirac operator only}, $\gamma^5\otimes D_R$, is 
  \begin{equation*}
    \label{eq:15}
    \A'_G \longrightarrow \A''_G = (\mathbb H_L
    \oplus \mathbb H'_L \oplus \C_R \oplus \C'_R) \oplus (\C_l
    \oplus M_3(\C)_l \oplus \C_r \oplus M_3(\C)_r) 
  \end{equation*}
with $\C_R = \C_r=\C_l.$
The main result of \cite{Devastato:2013fk} is that for $A\in\A''_G$
\begin{equation}
[\gamma^5\otimes D_R, A] \text{ is not necessarily zero.}\label{eq:11}
\end{equation}
In other terms, starting from the grand algebra one can generate the
field $\sigma$ by a fluctuation of the Majorana mass term which
respects the first order condition imposed by the Majorana mass term.

The further reduction to the standard model, that is 
  $\C'_R = \C_R, \mathbb H'_L = \mathbb H_L, M_3(\C)_l =M_3(\C)_r$,
  is obtained by the first order condition on the non-Majorana part of
  the Dirac operator.
\bigskip
  
Let us now discuss our result.
The representation \eqref{eq:17} of $C^\infty(\M)\otimes \A_G$
together with the Dirac operator in \eqref{eq:2bis} do not yield a
spectral triple, because whatever $A\in C^\infty(\M)\otimes \A_G$, the operator
\begin{equation}
  \label{eq:8}
  [\ds\otimes {\mathbb I}_{32}, A] = P + T^\mu \partial_\mu
\end{equation}
($P$ and $T^\mu$ are matrices whose explicit form is given in
\cite{Devastato:2013fk}) is not bounded. It becomes bounded
if $T^\mu =0$ for all $\mu=1,.., 4$, but this precisely means that the
action of the grand algebra should be diagonal on the spinorial
indices, meaning the reduction of $\A_G$ to $\A_{sm}$. Thus the
mixing of the spin and internal indices has two consequences: 

- the bounded bosonic operators~$B$ generated by a fluctuation
of the Dirac operator include the field $\sigma$; 

- there is
a ``deeper
alteration of spacetime'', encoded within the unbounded operator
$T^\mu\partial_\mu$. 

\noindent This last point is the most open to drastic changes  which may
be imagined, for example like dropping out the associativity of the algebra \cite{Farnsworth:2013vn}.

Alternatively, one may
imagine a cosmological scenario beginning with a ``pre-geometric
phase'', described by the grand algebra and the finite dimensional
Dirac operator $\gamma^5\otimes D_R$. The right neutrino would then
play the role of a ``primary elementary particle'', that generates the field $\sigma$. Then usual
geometry (encoded within the free Dirac operator $\ds\otimes\mathbb
I$) emerges at a later stage, and provokes the reduction to the SM. This makes and interesting echo to a recent inflationary
interpretation of the field $\sigma$ \cite{Bars:2013ys}. Moreover very recent data~\cite{BICEP} seem to indicate an inflationary scale at a scale of $10^{16}$~GeV, a scale in broad agreement with the unification of the coupling constant required by this approach. From a more mathematical point
of view, we stress that the triple $(C^\infty(\M)\otimes \A_G, \hh, \gamma^5\otimes D_R)$ satisfies the bounded commutator condition, but $\gamma^5\otimes D_R$
has no compact resolvent. 

To summarize, the grand algebra transfers
the problem of generating $\sigma$ from the noncommutative to the commutative 
part of the geometry: with the algebra of the
standard model, $C^\infty(\M)\otimes \A_{sm}$, the first
order condition is always satisfied by the free Dirac operator, the
problem is all in $D_R$.  Using
the grand algebra, we have that
$\gamma^5\otimes D_R$ both generates the field $\sigma$ and satisfies the first-order condition. But the free Dirac
does not satisfies this condition (neither the
bounded commutator one). Of course this is not satisfactory but
this suggests interesting path to explore.

Another question is whether the reduction to the SM imposed by the
first order condition can be understood dynamically (i.e. by a
minimization of the spectral action), as in the model of Chamseddine,
Connes and van Suijlekom where $\sigma$ is generated from $\A_{sm}$ 
by a fluctuation without first order
condition. 

\begin{acknowledgement}
 The authors thank W. van Suijlekom for discussion.
\end{acknowledgement}


\begin{thebibliography}{10}

\bibitem{BICEP}
P.A.R.Ade {\it et al.}  [ BICEP2 Collaboration],
  \emph{BICEP2 I: Detection Of B-mode Polarization at Degree Angular Scales,}
  arXiv:1403.3985 [astro-ph.CO];
  \emph{BICEP2 II: Experiment and Three-Year Data Set,}
  arXiv:1403.4302 [astro-ph.CO].


\bibitem{Bars:2013ys}
I.~Bars, P.~J. Steinhardt, and N.~Turok, \emph{Cyclic cosmology, conformal
  symmetry and the metastability of the higgs}, {P}reprint arXiv 1307.8106
  (2013).

\bibitem{Chamseddine:1996kx}
A.~H. Chamseddine and A.~Connes, \emph{The spectral action principle}, Commun.
  Math. Phys. \textbf{186} (1996), 737--750.

\bibitem{Chamseddine:2008uq}
A.~H. Chamseddine and A.~Connes, \emph{Why the standard model ?}, J. Geom. Phys \textbf{58} (2008),
  38--47.

\bibitem{Chamseddine:2012fk}
A.~H. Chamseddine and A.~Connes, \emph{Resilience of the spectral standard model}, JHEP \textbf{104}
  (2012).

\bibitem{Chamseddine:2007oz}
A.~H. Chamseddine, A.~Connes, and M.~Marcolli, \emph{Gravity and the standard
  model with neutrino mixing}, Adv. Theor. Math. Phys. \textbf{11} (2007),
  991--1089.

\bibitem{Chamseddine:2013uq}
A.~H. Chamseddine, A.~Connes, and Walter van Suijlekom, \emph{Beyond the
  spectral standard model: emergence of {P}ati-{S}alam unification}, JHEP
  \textbf{1311} (2013), 132.

\bibitem{Chamseddine:2013fk}
A.~H. Chamseddine, A.~Connes, and Walter van Suijlekom, \emph{{I}nner fluctuations in noncommutative geometry without first
  order condition}, J. Geom. Phy. \textbf{73} (2013).

\bibitem{Connes:1996fu}
A.~Connes, \emph{Gravity coupled with matter and the foundations of
  noncommutative geometry}, Commun. Math. Phys. \textbf{182} (1996), 155--176.

\bibitem{Connes:1990ix}
A.~Connes, J.~Lott, \emph{Particle models and noncommutative geo.},
  Nuclear Phys. B Proc. Suppl. \textbf{18B} (1990), 29--47.

\bibitem{connesreconstruct}
Alain Connes, \emph{On the spectral characterization of manifolds}, J. Noncom.
  Geom. \textbf{7} (2013), no.~1, 1--82.

\bibitem{Degrassi:2012fk}
G.~Degrassi, S.~Di Vita, J.~Elias-Miro, J.~R. Espinosa, G.~F. Giudice,
  G.~Isidori, and A.~Strumia, \emph{Higgs mass and vacuum stability in the
  standard model at nnlo}, JHEP {\bf 08} (2012) 098.

\bibitem{Devastato:2013fk}
A.~Devastato, F.~Lizzi, P.~Martinetti, \emph{{G}rand {S}ymmetry, {S}pectral
  {A}ction and the {H}iggs mass}, JHEP {\bf 01}~(2014)~042.

\bibitem{Dubois-Violette:1989fk}
M.~Dubois-Violette, J.~Madore, and R.~Kerner, \emph{Classical bosons in a
  noncommutative geometry}, Class. Quantum Grav. \textbf{1709} (1989).

\bibitem{Elias-Miro:2012ys}
J.~Elias-Miro, J.~R. Espinosa, G.~F. Giudice, H.~M. Lee, and A.~Strumia.
\emph{Stabilization of the electroweak vacuum by a scalar threshold effect},
 JHEP {\bf 06} (2012) 031.

\bibitem{Farnsworth:2013vn}
S.~Farnsworth and L.~Boyle, \emph{Non-associative geometry and the spectral
  action principle}, {P}reprint arXiv 1303.1782 (2013).

\bibitem{Hertzberg:2012fk}
M.~P. Hertzberg, \emph{A correlation between the higgs mass and dark matter},
  (2012), arXiv: 1210.3624.
  
 \bibitem{fermiondoubling} F.~Lizzi, G.~Mangano, G.~Miele and G.~Sparano,
 \emph{Fermion Hilbert space and fermion doubling in the noncommutative geometry approach to gauge theories,}
  Phys.\ Rev.\ D {\bf 55} (1997) 6357
  [hep-th/9610035].


\bibitem{Stephan:2009fk}
C.~A. Stephan, \emph{New scalar fields in noncommutative geometry}, Phys. Rev.
  D \textbf{79} (2009).

\bibitem{schuckhiggs}
T. Sch\"ucker, \emph{Higgs mass predictions}, arXiv 0708.3344 [hep-th] 

\end{thebibliography}

\end{document}